# New Physics of Metamaterials

Zhong-Yue Wang[1]

*Engineering Department, SBSC, No.580 Songhuang Road, Shanghai 201703, China*

**Abstract:** Einstein utilized Lorentz invariance from Maxwell's equations to modify mechanical laws and establish the special theory of relativity. Similarly, we may have a different theory if there exists another covariance of Maxwell's equations. In this paper, we find such a new transformation where Maxwell's equations are still unchanged. Consequently, Veselago's metamaterial and other systems have negative phase velocities without double negative permittivity and permeability can be described by a unified theory. People are interested in the application of metamaterials and negative phase velocities but do not appreciate the magnitude and significance to the spacetime conception of modern physics and philosophy.

1.*Corresponding author, E-mail: zhongyuewang (at) ymail.com*

# I..... Introduction

Consider an inertial reference frame $K'$ moves at a constant velocity $V$ with respect to another inertial system $K$ as shown in Fig.1.

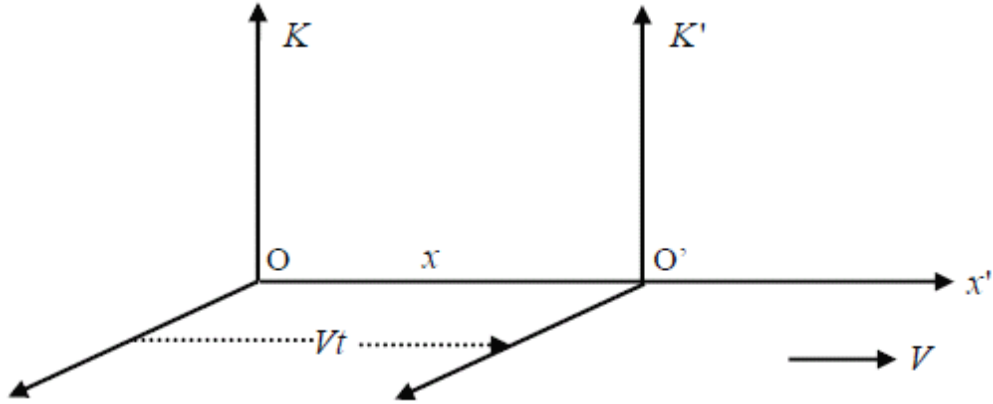

Figure 1.  Two inertial frames with a common axis

For convenience, the three sets of axes are parallel and their relative motion is along the common $x$-$x'$ axis. The form of Maxwell's equations in $K'$ and $K$ does not change under the following Lorentz transformation

$$x' = \frac{x - Vt}{\sqrt{1 - \frac{V^2}{C^2}}} \tag{1}$$

$$y' = y \tag{2}$$

$$z' = z \tag{3}$$

$$t' = \frac{t - \frac{V}{C^2}x}{\sqrt{1 - \frac{V^2}{C^2}}} \tag{4}$$

$$C^2 = \frac{1}{\varepsilon \mu} \tag{5}$$

Einstein assumed mechanical laws to satisfy this transformation and get

$$\mathbf{p} = m\mathbf{V} \xrightarrow{V \ll C} m_0 \mathbf{V} \quad (m = \frac{m_0}{\sqrt{1 - \frac{V^2}{C^2}}}) \tag{6}$$

$$E = mC^2 \tag{7}$$

$$E^2 = p^2 C^2 + m_0^2 C^4 \tag{8}$$

Although the Lorentz transformation is seemingly strange and complex, it can reduce to the Galilean transformation at low speeds ($V \ll C$)

$$x' = x - Vt \tag{9}$$

$$y' = y \tag{10}$$

$$z' = z \tag{11}$$

$$t' = t \tag{12}$$

which agrees with our classical intuition. Addition of velocities is

$$\frac{dx'}{dt'} = \frac{dx}{dt} - V \tag{13}$$

However, nobody had ever though of the following space-time transformation to Fig.1,

$$x' = \frac{x + Vt}{\sqrt{1 - \frac{V^2}{C^2}}} \tag{14}$$

$$y' = y \tag{15}$$

$$z' = z \tag{16}$$

$$t' = \frac{t + \frac{V}{C^2} x}{\sqrt{1 - \frac{V^2}{C^2}}} \tag{17}$$

because the consequence is inconsistent with common senses even in the classical limit. For example,

$$x' = \frac{x+Vt}{\sqrt{1-\frac{V^2}{C^2}}} \xrightarrow{V \ll C} x+Vt \qquad (18)$$

$$\frac{dx'}{dt'} = \frac{\frac{dx}{dt}+V}{1+\frac{V}{C^2}\frac{dx}{dt}} \xrightarrow{V \ll C} \frac{dx}{dt}+V \qquad (19)$$

Hereinafter, we show that Maxwell's equations are invariant under the incredible transformation (14)-(17) and develop a tentative mechanical theory to compare with experiments.

## II. Anomalous Invariance and Maxwell's Equations

Maxwell's equations in $K$ are

$$\frac{\partial E_x}{\partial x}+\frac{\partial E_y}{\partial y}+\frac{\partial E_z}{\partial z} = \frac{\rho}{\varepsilon} \qquad (20)$$

$$\frac{\partial E_z}{\partial y}-\frac{\partial E_y}{\partial z} = -\frac{\partial B_x}{\partial t} \qquad (21\text{-}x)$$

$$\frac{\partial E_x}{\partial z}-\frac{\partial E_z}{\partial x} = -\frac{\partial B_y}{\partial t} \qquad (21\text{-}y)$$

$$\frac{\partial E_y}{\partial x}-\frac{\partial E_x}{\partial y} = -\frac{\partial B_z}{\partial t} \qquad (21\text{-}z)$$

$$\frac{\partial B_x}{\partial x}+\frac{\partial B_y}{\partial y}+\frac{\partial B_z}{\partial z} = 0 \qquad (22)$$

$$\frac{\partial B_z}{\partial y}-\frac{\partial B_y}{\partial z} = \mu j_x + \frac{1}{C^2}\frac{\partial E_x}{\partial t} \qquad (23\text{-}x)$$

$$\frac{\partial B_x}{\partial z} - \frac{\partial B_z}{\partial x} = \mu j_y + \frac{1}{C^2}\frac{\partial E_y}{\partial t} \tag{23-y}$$

$$\frac{\partial B_y}{\partial x} - \frac{\partial B_x}{\partial y} = \mu j_z + \frac{1}{C^2}\frac{\partial E_z}{\partial t} \tag{23-z}$$

According to (14)-(17),

$$\frac{\partial}{\partial x} = \frac{1}{\sqrt{1-\frac{V^2}{C^2}}}\left(\frac{\partial}{\partial x'} + \frac{V}{C^2}\frac{\partial}{\partial t'}\right) \tag{24}$$

$$\frac{\partial}{\partial y} = \frac{\partial}{\partial y'} \tag{25}$$

$$\frac{\partial}{\partial z} = \frac{\partial}{\partial z'} \tag{26}$$

$$\frac{\partial}{\partial t} = \frac{1}{\sqrt{1-\frac{V^2}{C^2}}}\left(\frac{\partial}{\partial t'} + V\frac{\partial}{\partial x'}\right) \tag{27}$$

Physical quantities in $K'$,

$$E_x' = E_x \qquad E_y' = \frac{E_y + VB_z}{\sqrt{1-\frac{V^2}{C^2}}} \qquad E_z' = \frac{E_z - VB_y}{\sqrt{1-\frac{V^2}{C^2}}} \tag{28}$$

$$B_x' = B_x \qquad B_y' = \frac{B_y - \frac{V}{C^2}E_z}{\sqrt{1-\frac{V^2}{C^2}}} \qquad B_z' = \frac{B_z + \frac{V}{C^2}E_y}{\sqrt{1-\frac{V^2}{C^2}}} \tag{29}$$

$$\rho' = \frac{\rho + \frac{V}{C^2}j_x}{\sqrt{1-\frac{V^2}{C^2}}} \qquad j_x' = \frac{j_x + \rho V}{\sqrt{1-\frac{V^2}{C^2}}} \qquad j_y' = j_y \qquad j_z' = j_z \tag{30}$$

In $K$,

$$E_x = E_x' \tag{31}$$

$$E_y = \frac{E_y' - VB_z'}{\sqrt{1 - \frac{V^2}{C^2}}} \tag{32}$$

$$E_z = \frac{E_z' + VB_y'}{\sqrt{1 - \frac{V^2}{C^2}}} \tag{33}$$

$$B_x = B_x' \tag{34}$$

$$B_y = \frac{B_y' + \frac{V}{C^2} E_z'}{\sqrt{1 - \frac{V^2}{C^2}}} \tag{35}$$

$$B_z = \frac{B_z' - \frac{V}{C^2} E_y'}{\sqrt{1 - \frac{V^2}{C^2}}} \tag{36}$$

$$\rho = \frac{\rho' - \frac{V}{C^2} j_x'}{\sqrt{1 - \frac{V^2}{C^2}}} \tag{37}$$

$$j_x = \frac{j_x' - \rho' V}{\sqrt{1 - \frac{V^2}{C^2}}} \tag{38}$$

$$j_y = j_y' \tag{39}$$

$$j_z = j_z' \tag{40}$$

Substituting Eqs.(24)-(26),(31)-(33),(37) and (5) into (20),

$$\frac{1}{\sqrt{1-\frac{V^2}{C^2}}}\left(\frac{\partial}{\partial x'}+\frac{V}{C^2}\frac{\partial}{\partial t'}\right)E_x'+\frac{\partial}{\partial y'}\frac{E_y'-VB_z'}{\sqrt{1-\frac{V^2}{C^2}}}+\frac{\partial}{\partial z'}\frac{E_z'+VB_y'}{\sqrt{1-\frac{V^2}{C^2}}}=\frac{\rho'-\frac{V}{C^2}j_x'}{\varepsilon\sqrt{1-\frac{V^2}{C^2}}} \quad (41)$$

$$\frac{\partial E_x'}{\partial x'}+\frac{\partial E_y'}{\partial y'}+\frac{\partial E_z'}{\partial z'}-V\left(\frac{\partial B_z'}{\partial y'}-\frac{\partial B_y'}{\partial z'}-\frac{1}{C^2}\frac{\partial E_x'}{\partial t'}\right)=\frac{\rho'}{\varepsilon}-V\mu j_x' \quad (42)$$

i.e.
$$\frac{\partial E_x'}{\partial x'}+\frac{\partial E_y'}{\partial y'}+\frac{\partial E_z'}{\partial z'}=\frac{\rho'}{\varepsilon} \quad (20')$$

$$\frac{\partial B_z'}{\partial y'}-\frac{\partial B_y'}{\partial z'}=\mu j_x'+\frac{1}{C^2}\frac{\partial E_x'}{\partial t'} \quad (23\text{-x'})$$

Likewise, (21-x) is now

$$\frac{\partial}{\partial y'}\frac{E_z'+VB_y'}{\sqrt{1-\frac{V^2}{C^2}}}-\frac{\partial}{\partial z'}\frac{E_y'-VB_z'}{\sqrt{1-\frac{V^2}{C^2}}}=-\frac{1}{\sqrt{1-\frac{V^2}{C^2}}}\left(\frac{\partial}{\partial t'}+V\frac{\partial}{\partial x'}\right)B_x' \quad (43)$$

$$\frac{\partial E_z'}{\partial y'}-\frac{\partial E_y'}{\partial z'}+\frac{\partial B_x'}{\partial t'}+V\left(\frac{\partial B_x'}{\partial x'}+\frac{\partial B_y'}{\partial y'}+\frac{\partial B_z'}{\partial z'}\right)=0 \quad (44)$$

i.e.
$$\frac{\partial E_z'}{\partial y'}-\frac{\partial E_y'}{\partial z'}=-\frac{\partial B_x'}{\partial t'} \quad (21\text{-x'})$$

$$\frac{\partial B_x'}{\partial x'}+\frac{\partial B_y'}{\partial y'}+\frac{\partial B_z'}{\partial z'}=0 \quad (22')$$

Eq(21-y):

$$\frac{\partial E_x'}{\partial z'} - \frac{1}{\sqrt{1-\frac{V^2}{C^2}}}\left(\frac{\partial}{\partial x'} + \frac{V}{C^2}\frac{\partial}{\partial t'}\right)\frac{E_z'+VB_y'}{\sqrt{1-\frac{V^2}{C^2}}} = -\frac{1}{\sqrt{1-\frac{V^2}{C^2}}}\left(\frac{\partial}{\partial t'} + V\frac{\partial}{\partial x'}\right)\frac{B_y'+\frac{V}{C^2}E_z'}{\sqrt{1-\frac{V^2}{C^2}}} \quad (45)$$

$$\frac{\partial E_x'}{\partial z'} - \frac{1}{1-\frac{V^2}{C^2}}\left(1-\frac{V^2}{C^2}\right)\frac{\partial E_z'}{\partial x'} = -\frac{1}{1-\frac{V^2}{C^2}}\left(1-\frac{V^2}{C^2}\right)\frac{\partial B_y'}{\partial t'} \quad (46)$$

$$\frac{\partial E_x'}{\partial z'} - \frac{\partial E_z'}{\partial x'} = -\frac{\partial B_y'}{\partial t'} \quad (21\text{-y'})$$

Eq(21-z):

$$\frac{1}{\sqrt{1-\frac{V^2}{C^2}}}\left(\frac{\partial}{\partial x'} + \frac{V}{C^2}\frac{\partial}{\partial t'}\right)\frac{E_y'-VB_z'}{\sqrt{1-\frac{V^2}{C^2}}} - \frac{\partial}{\partial y'}E_x' = -\frac{1}{\sqrt{1-\frac{V^2}{C^2}}}\left(\frac{\partial}{\partial t'} + V\frac{\partial}{\partial x'}\right)\frac{B_z'-\frac{V}{C^2}E_y'}{\sqrt{1-\frac{V^2}{C^2}}} \quad (47)$$

$$\frac{1}{1-\frac{V^2}{C^2}}\left(1-\frac{V^2}{C^2}\right)\frac{\partial E_y'}{\partial x'} - \frac{\partial E_x'}{\partial y'} = -\frac{1}{1-\frac{V^2}{C^2}}\left(1-\frac{V^2}{C^2}\right)\frac{\partial B_z'}{\partial t'} \quad (48)$$

$$\frac{\partial E_y'}{\partial x'} - \frac{\partial E_x'}{\partial y'} = -\frac{\partial B_z'}{\partial t'} \quad (21\text{-z'})$$

Eq(22):

$$\frac{1}{\sqrt{1-\frac{V^2}{C^2}}}\left(\frac{\partial}{\partial x'} + \frac{V}{C^2}\frac{\partial}{\partial t'}\right)B_x' + \frac{\partial}{\partial y'}\frac{B_y'+\frac{V}{C^2}E_z'}{\sqrt{1-\frac{V^2}{C^2}}} + \frac{\partial}{\partial z'}\frac{B_z'-\frac{V}{C^2}E_y'}{\sqrt{1-\frac{V^2}{C^2}}} = 0 \quad (49)$$

$$\frac{\partial B_x'}{\partial x'} + \frac{\partial B_y'}{\partial y'} + \frac{\partial B_z'}{\partial z'} + \frac{V}{C^2}\left(\frac{\partial E_z'}{\partial y'} - \frac{\partial E_y'}{\partial z'} + \frac{\partial B_x'}{\partial t'}\right) = 0 \quad (50)$$

i.e.

$$\frac{\partial B_x'}{\partial x'}+\frac{\partial B_y'}{\partial y'}+\frac{\partial B_z'}{\partial z'}=0 \qquad (22')$$

$$\frac{\partial E_z'}{\partial y'}-\frac{\partial E_y'}{\partial z'}=-\frac{\partial B_x'}{\partial t'} \qquad (21\text{-}x')$$

Eq(23-x):

$$\frac{\partial}{\partial y'}\frac{B_z'-\frac{V}{C^2}E_y'}{\sqrt{1-\frac{V^2}{C^2}}}-\frac{\partial}{\partial z'}\frac{B_y'+\frac{V}{C^2}E_z'}{\sqrt{1-\frac{V^2}{C^2}}}=\mu\frac{j_x'-\rho'V}{\sqrt{1-\frac{V^2}{C^2}}}+\frac{1}{C^2\sqrt{1-\frac{V^2}{C^2}}}\left(\frac{\partial}{\partial t'}+V\frac{\partial}{\partial x'}\right)E_x' \qquad (51)$$

$$\frac{\partial B_z'}{\partial y'}-\frac{\partial B_y'}{\partial z'}=\mu j_x'-\mu\rho'V+\frac{1}{C^2}\frac{\partial E_x'}{\partial t'}+\frac{V}{C^2}\left(\frac{\partial E_x'}{\partial x'}+\frac{\partial E_y'}{\partial y'}+\frac{\partial E_z'}{\partial z'}\right) \qquad (52)$$

Owing to $\frac{1}{C^2}=\varepsilon\mu$ (5) and $\frac{\partial E_x'}{\partial x'}+\frac{\partial E_y'}{\partial y'}+\frac{\partial E_z'}{\partial z'}=\frac{\rho'}{\varepsilon}$ (20'), Eq.(52) is

$$\frac{\partial B_z'}{\partial y'}-\frac{\partial B_y'}{\partial z'}=\mu j_x'+\frac{1}{C^2}\frac{\partial E_x'}{\partial t'} \qquad (23\text{-}x')$$

Eq(23-y):

$$\frac{\partial B_x'}{\partial z'}-\frac{1}{\sqrt{1-\frac{V^2}{C^2}}}\left(\frac{\partial}{\partial x'}+\frac{V}{C^2}\frac{\partial}{\partial t'}\right)\frac{B_z'-\frac{V}{C^2}E_y'}{\sqrt{1-\frac{V^2}{C^2}}}=\mu j_y'+\frac{1}{C^2\sqrt{1-\frac{V^2}{C^2}}}\left(\frac{\partial}{\partial t'}+V\frac{\partial}{\partial x'}\right)\frac{E_y'-VB_z'}{\sqrt{1-\frac{V^2}{C^2}}} \qquad (53)$$

$$\frac{\partial B_x'}{\partial z'}-\frac{1}{1-\frac{V^2}{C^2}}\left(1-\frac{V^2}{C^2}\right)\frac{\partial B_z'}{\partial x'}=\mu j_y'+\frac{1}{C^2\left(1-\frac{V^2}{C^2}\right)}\left(1-\frac{V^2}{C^2}\right)\frac{\partial E_y'}{\partial t'} \qquad (54)$$

$$\frac{\partial B_x'}{\partial z'}-\frac{\partial B_z'}{\partial x'}=\mu j_y'+\frac{1}{C^2}\frac{\partial E_y'}{\partial t'} \qquad (23\text{-}y')$$

Eq(23-z):

$$\frac{1}{\sqrt{1-\frac{V^2}{C^2}}}\left(\frac{\partial}{\partial x'}+\frac{V}{C^2}\frac{\partial}{\partial t'}\right)\frac{B_y'+\frac{V}{C^2}E_z'}{\sqrt{1-\frac{V^2}{C^2}}}-\frac{\partial B_x'}{\partial y'}=\mu j_z'+\frac{1}{C^2\sqrt{1-\frac{V^2}{C^2}}}\left(\frac{\partial}{\partial t'}+V\frac{\partial}{\partial x'}\right)\frac{E_z'+VB_y'}{\sqrt{1-\frac{V^2}{C^2}}} \quad (55)$$

$$\frac{1}{1-\frac{V^2}{C^2}}\left(1-\frac{V^2}{C^2}\right)\frac{\partial B_y'}{\partial x'}-\frac{\partial B_x'}{\partial y'}=\mu j_z'+\frac{1}{C^2\left(1-\frac{V^2}{C^2}\right)}\left(1-\frac{V^2}{C^2}\right)\frac{\partial E_z'}{\partial t'} \quad (56)$$

$$\frac{\partial B_y'}{\partial x'}-\frac{\partial B_x'}{\partial y'}=\mu j_z'+\frac{1}{C^2}\frac{\partial E_z'}{\partial t'} \quad (23\text{-}z')$$

Maxwell's equations remain in the same form under two transformations (1)-(4) and (14)-(17).

It is a pity that the latter had never been studied.

### III..... New Mechanics and Measurable Effects

Under this new transformation, the product of a physical quantity times the velocity $\mathbf{V}$ is reversed, e.g.

$$\mathbf{j}=-\rho\mathbf{V}, \quad \text{Vector Potential} \quad \mathbf{A}=-\frac{\phi}{C^2}\mathbf{V} \quad (57)$$

Introduce a mechanical theory which is consistent with (14)-(17), as Einstein had done to special relativity and the Lorentz transformation (1)-(4). Hence, the momentum and total energy should be

$$\mathbf{p}=-m\mathbf{V}\xrightarrow{V\ll C}-m_0\mathbf{V} \quad (m=\frac{m_0}{\sqrt{1-\frac{V^2}{C^2}}}) \quad (58)$$

$$E=mC^2 \quad (7)$$

$$E^2=p^2C^2+m_0^2C^4 \quad (8)$$

It is surprising that the momentum $\mathbf{p}$ is in a direction opposite to the velocity $\mathbf{V}$. That is to say, this new theory is symmetric to relativistic mechanics and Newtonian. In view of de Broglie's relation $\mathbf{p}=\hbar\mathbf{k}$, the wave vector $\mathbf{k}$ of a photon should be antiparallel to the arrival of the mass $m = \dfrac{\hbar\omega}{C^2}$ and energy $\hbar\omega$ (mass-energy equivalence). As electromagnetic waves consist of a stream of photons, the phase velocity $\dfrac{\omega}{k^2}\mathbf{k}$ [1] is also negative.

We can obtain Snell's law of refraction from conservation of the momentum component parallel to the interface for a single photon[2](Fig.2).

$$\hbar\mathbf{k}_1 \sin\gamma_1 = \hbar\mathbf{k}_2 \sin\gamma_2 \tag{59}$$

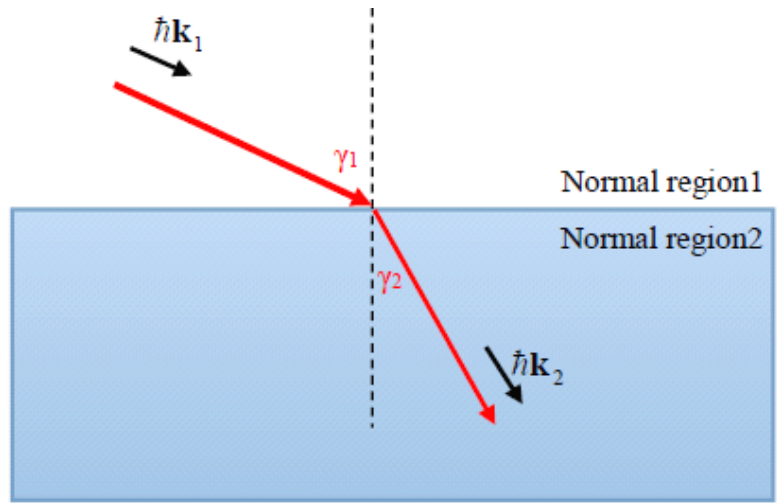

Figure 2.  Snell's law

If the momentum of a photon is opposite to the arrival of energy(red arrow), the light ray should be refracted on the same side of the incident beam otherwise the horizontal momentum of the photon is non-conservational (Fig.3).

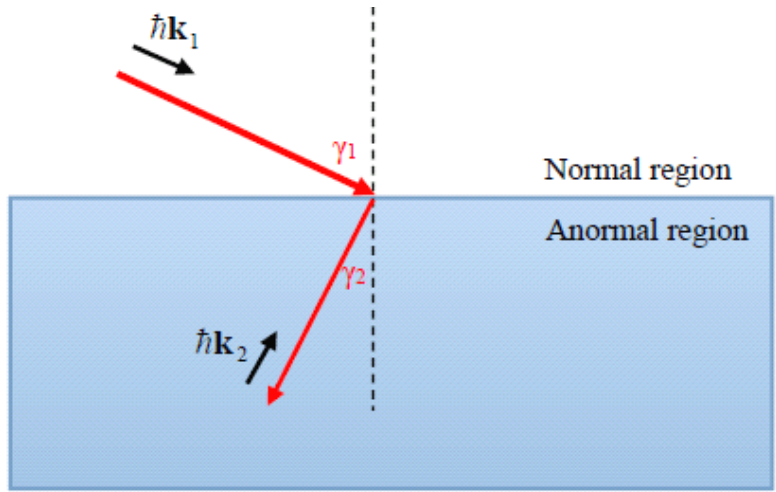

Figure 3.  Negative refraction

Moreover, the Cherenkov effect can be deduced from the photon theory[3][4] and energy-momentum conservation(Fig.4).

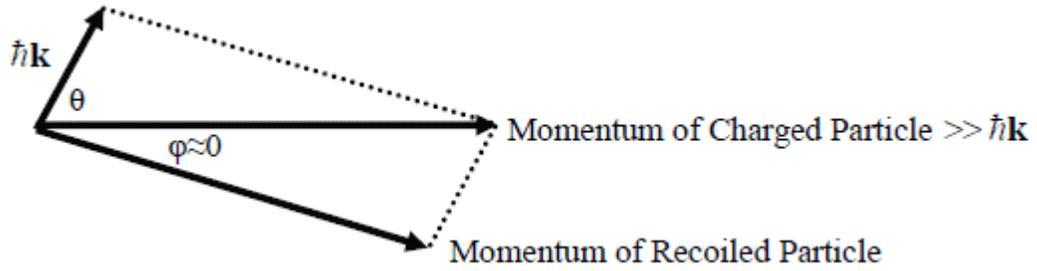

Figure 4.　Momentum Conservation

For normal photons(Fig.5),

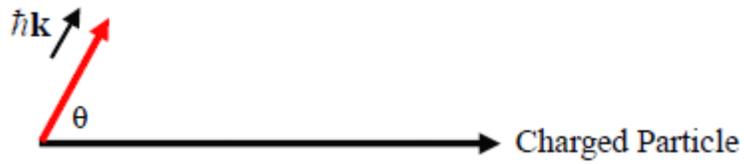

Figure 5.　Cherenkov radiation

As to photons have a negative momentum(Fig.6),

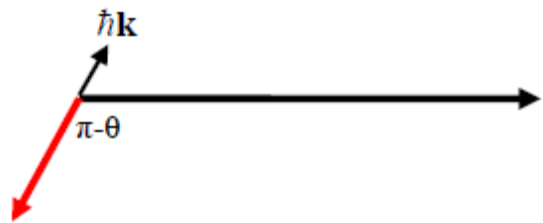

Figure 6.　Reversed Cherenkov radiation

Regular Doppler effect derived by Lorentz transformation:

$$\omega' = \frac{\omega - kV}{\sqrt{1-\frac{V^2}{C^2}}} < \omega \qquad \text{(Fig.7)} \qquad (60)$$

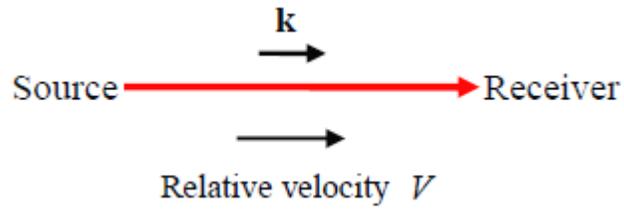

Figure 7.  Doppler effect (the receiver is moving away from the source)

$$\omega' = \frac{\omega + kV}{\sqrt{1-\frac{V^2}{C^2}}} > \omega \qquad \text{(Fig.8)} \qquad (61)$$

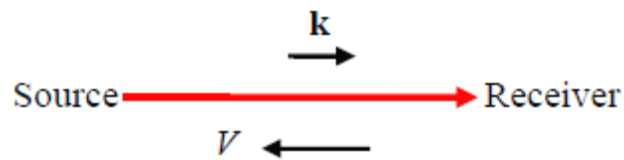

Figure 8.  Doppler effect (one approaches the other)

Inverse Doppler effect:

$$\omega' = \frac{\omega + kV}{\sqrt{1 - \frac{V^2}{C^2}}} > \omega \quad \text{(Fig.9)} \tag{62}$$

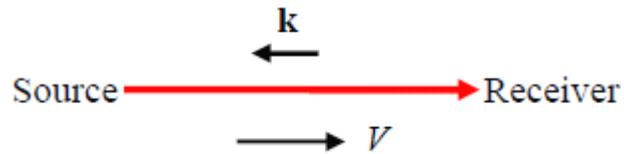

Figure 9. Inverse Doppler effect (they move away from each other)

$$\omega' = \frac{\omega - kV}{\sqrt{1 - \frac{V^2}{C^2}}} < \omega \quad \text{(Fig.10)} \tag{63}$$

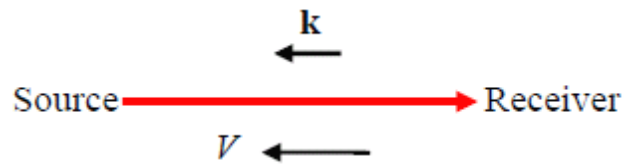

Figure 10. Inverse Doppler effect (one is moving towards the other)

## IV. Application: Veselago's Materials and Others

A method to realize (58) is to generate an even number of imaginary momenta and the final effective momentum is a negative quantity. For instance, single negative materials with $\varepsilon_{eff} < 0$ or $\mu_{eff} < 0$ cause an imaginary wave vector(momentum) respectively and the combination results in a negative wave vector(momentum). This is just the mechanism of Veselago's material[5] to have negative refraction[6], reversed Cherenkov effect[7] and inverse Doppler effect[8]. But $\varepsilon_{eff} < 0$ and $\mu_{eff} < 0$ are not necessary conditions. Actually, negative phase velocities sporadically appeared in other studies[9]~[12]. A practical structure is the slow wave device in vacuum electronics where the phase constant $\beta$ and phase velocity $\frac{\omega}{\beta}$ can be negative[13]. They have nothing to do with Veselago's proposal $\varepsilon_{eff} < 0$ and $\mu_{eff} < 0$.

## V. Physical Meaning of Advanced Potentials

Potentials with the time dependence $t - \frac{r}{\frac{1}{\sqrt{\varepsilon\mu}}}$ and $t + \frac{r}{\frac{1}{\sqrt{\varepsilon\mu}}}$ are called retarded potentials and advanced potentials. At present, advanced potentials are deemed to violate the principle of causality and are discarded although they are entirely consistent with Maxwell's equations[14]. In fact, $t + \frac{r}{\frac{1}{\sqrt{\varepsilon\mu}}}$ can be rewritten as $t - \frac{r}{-\frac{1}{\sqrt{\varepsilon\mu}}}$ which implies that the phase velocity is $V_p = -\frac{1}{\sqrt{\varepsilon\mu}} < 0$. "Advanced potentials" as potentials of negative phase velocities are not in conflict with causality.

## VI. Energy Transport and Poynting Vector

The energy $\hbar\omega$ and mass $m = \dfrac{\hbar\omega}{C^2}$ propagates from the source to the receiver, while the momentum $\mathbf{p} = -m\mathbf{V}$ is negative in this new theory(Fig.11).

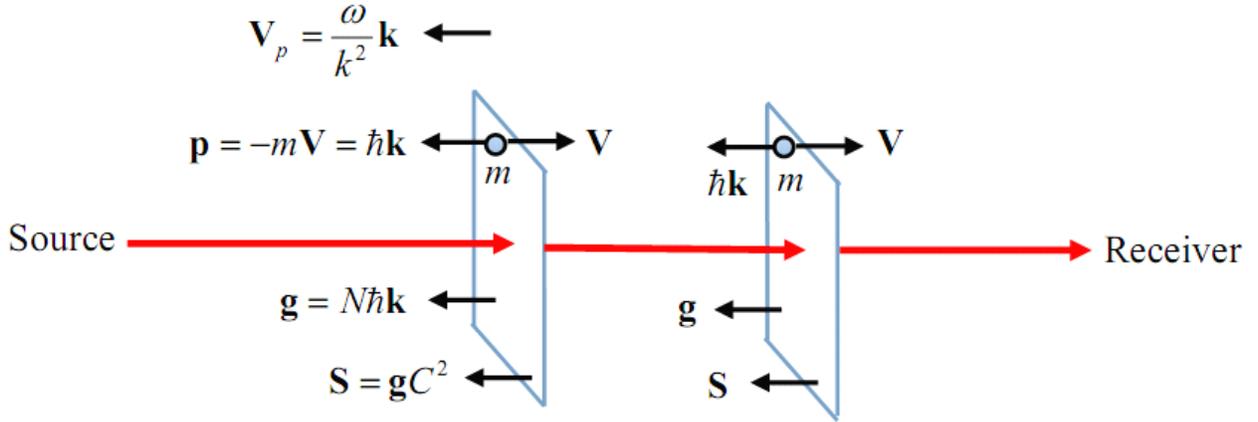

Figure 11. Energy(mass) transfer

The momentum density of an EM field

$$\mathbf{g} = N\mathbf{p} = -Nm\mathbf{V} = N\hbar\mathbf{k} \qquad (N \text{ number density of photons}) \qquad (64)$$

is in the direction of $\mathbf{k}$. In addition,

$$\mathbf{g}C^2 = -NmC^2\mathbf{V} = -N\hbar\omega\mathbf{V} = -w\mathbf{V} \qquad (w = N\hbar\omega > 0 \text{ density of energy}) \quad (65)$$

Therefore, the energy flux density defined as

$$\mathbf{S} = -w\mathbf{V} \qquad (66)$$

like (57) and (58) is equal to

$$\mathbf{g}C^2 = N\hbar\mathbf{k}C^2 \propto \mathbf{k} \qquad (67)$$

The Poynting vector $\mathbf{S}$ should be in the same direction of momentum density $\mathbf{g}$ and wave vector $\mathbf{k}$ (Fig.11).

# VII. Faster Than Light

The form of Eq.(7) is still tenable in a theory of superluminal bodies[15] and the criterion for a particle or wave to exceed $c = 299,752,498 \; m/s$ should be

$$V^2 > c^2 \tag{68}$$

$$\frac{p^2}{m^2} > c^2 \tag{69}$$

$$\frac{\hbar^2 k^2}{\left(\frac{\hbar \omega}{C^2}\right)^2} > c^2 \tag{70}$$

$$k^2 > \omega^2 \frac{c^2}{C^4} \tag{71}$$

$$k^2 > \frac{\omega^2}{c^2} \quad (C = c) \tag{72}$$

The dispersion relation

$$\varepsilon_{eff} = \varepsilon_0 \left( 1 - \frac{\omega_p^2}{\omega^2 - \omega_0^2} \right) \tag{73}$$

$$\mu_{eff} = \mu_0 \left( 1 - \frac{F\omega^2}{\omega^2 - \omega_m^2} \right) \tag{74}$$

$$k^2 = \omega^2 \varepsilon_{eff} \mu_{eff} = \frac{\omega^2}{c^2} \left( 1 - \frac{\omega_p^2}{\omega^2 - \omega_0^2} \right) \left( 1 - \frac{F\omega^2}{\omega^2 - \omega_m^2} \right) \tag{75}$$

can meet Eq.(72) in the range $\omega < \frac{\omega_m}{\sqrt{1-F}}$, $\omega_0 < \omega \ll \omega_p$ (double negative) or $\omega < \omega_m$, $\omega < \omega_0$ (double positive). We plan to design suitable materials to observe[16]. The angular frequency $\omega$ in a double positive material cannot be much less than $\omega_0$ otherwise

$$\varepsilon_{eff} \approx \varepsilon_0 \left( 1 + \frac{\omega_p^2}{\omega_0^2} \right) = cons \tan t \tag{76}$$

This behaves as if a dielectric medium where $c$ is replaced by $C$ [17][18] and Eq.(71) should be applied. It makes against our aim to break the light barrier.

Sometimes the momentum of a photon is $\mathbf{p} = \hbar\boldsymbol{\beta}$ [4] and the criterion should now be

$$\frac{\hbar^2 \beta^2}{\left(\dfrac{\hbar\omega}{C^2}\right)^2} > c^2 \tag{77}$$

$$\beta^2 > \omega^2 \frac{c^2}{C^4} \tag{78}$$

In vacuum,

$$\beta^2 > \frac{\omega^2}{c^2} \tag{79}$$

$$\frac{\omega^2}{\beta^2} < c^2 \tag{80}$$

describes slow phase velocity waves[13].

## Conclusion

We construct a new system from first principles to replace Veselago's phenomenological theory. It can interpret all kinds of negative phase velocities which essentially represent a momentum antiparallel to the velocity. Retarded potentials and advanced potentials correspond to Lorentz invariance and the anomalous space-time transformation respectively. Faster than light energy is possible to exist in a metamaterial.

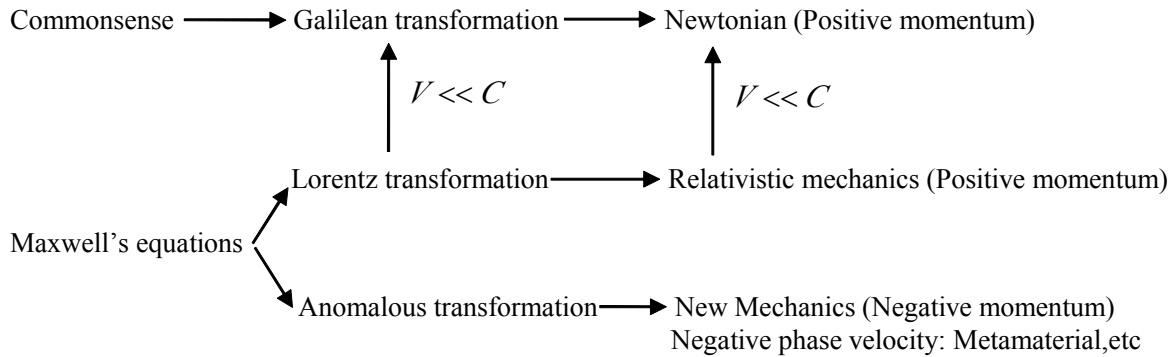

Commonsense ⟶ Galilean transformation ⟶ Newtonian (Positive momentum)

↑ $V \ll C$       ↑ $V \ll C$

Lorentz transformation ⟶ Relativistic mechanics (Positive momentum)

Maxwell's equations

Anomalous transformation ⟶ New Mechanics (Negative momentum)
Negative phase velocity: Metamaterial, etc

# References

[1]. E.H.Wichmann, ***Quantum Physics,*** Berkeley Physics Course Vol4, McGrow Hill, p.182
[2]. E.Hecht, ***Optics***, 4th edition, Addison-Wesley, p141
[3]. Z.Y.Wang, "Graphene, neutrino mass and oscillation", arXiv:0909.1856v2
[4]. Z.Y.Wang, "Extension to de Broglie formula and momentum operator of Quantum Mechanics", to be published
[5]. V.G.Veselago, "The electrodynamics of substances with simultaneously negative values of $\varepsilon$ and $\mu$", ***Sov. Phys. Usp.*** **10**(4), 509-514(1968)
[6]. R.A.Shelby, D.R.Smith and S.Schultz, "Experimental verification of a negative index of refraction", ***Science,*** **292**(5514), 77-79(2001)
[7]. S.Xi, H.S.Chen, T.Jiang, *et al,* "Experimental Verification of Reversed Cherenkov Radiation in Left-Handed Metamaterial", ***Phys. Rev. Lett.,*** **103**(19): 194801(2009)
[8]. N.Seddon and T.Bearpark, "Observation of the inverse Doppler effect", ***Science,*** **302**(5650), 1537-1540(2003)
[9]. Z.X.Huang, ***Introduction to the Theory of Cut-off Waveguides***(in Chinese), Metrology Publishing, 1991, p.145
[10]. K.Wynne, J.Carey, J.Zawadzka, *et al,* "Tunneling of single-cycle terahertz pulses through waveguides", ***Opt.Commun,*** **176**(4-6), 429-435(2000)
[11]. C.Luo, M.Ibanescu, S.G.Johnson, *et al,* "Cerenkov Radiation in Photonic Crystals", ***Science,*** **299**(5605), 368-371(2003)
[12]. Y. Zhang, B. Fluegel and A. Mascarenhas, "Total Negative Refraction in Real Crystals for Ballistic Electrons and Light", ***Phys. Rev. Lett.,*** **91**(15): 157404(2003)
[13]. K.Zhang and D.Li, ***Electromagnetic theory for Microwaves and Optoelectronics***,2nd edition, Springer, pp.401-472
[14]. D.J.Griffiths, ***Introduction to Electrodynamics,*** 3rd edition, Prentice Hall, p.425
[15]. G.Feinberg, "Possibility of Faster-Than-Light Particles", ***Phys. Rev.*** **159**, 1089–1105 (1967)
[16]. J.Fan and Z.Y.Wang, "New method to measure energy velocity and experimental test in dielectric media", to be published
[17]. Z.Y.Wang, P.Y.Wang and Y.R.Xu, "Crucial experiment to resolve Abraham–Minkowski controversy", ***Opnk.*** **122** (22), 1994-1996(2011)
[18]. Z.Y.Wang, "Relativistic Electrodynamics in media and generalized Lorentz transformation", to be published

## Postscript

Equations

$$\nabla \cdot \mathbf{E} = -\frac{\rho}{\varepsilon} \tag{81}$$

$$\nabla \times \mathbf{E} = -\frac{\partial}{\partial t}\mathbf{B} \tag{82}$$

$$\nabla \cdot \mathbf{B} = 0 \tag{83}$$

$$\nabla \times \mathbf{B} = \mu\mathbf{j} - \varepsilon\mu\frac{\partial}{\partial t}\mathbf{E} \tag{84}$$

and

$$\nabla \cdot \mathbf{E} = \frac{\rho}{\varepsilon} \tag{85}$$

$$\nabla \times \mathbf{E} = -\frac{\partial}{\partial t}\mathbf{B} \tag{86}$$

$$\nabla \cdot \mathbf{B} = 0 \tag{87}$$

$$\nabla \times \mathbf{B} = -\mu\mathbf{j} - \varepsilon\mu\frac{\partial}{\partial t}\mathbf{E} \tag{88}$$

are invariant under

$$x' = \frac{x - Vt}{\sqrt{1 + \frac{V^2}{C^2}}} \tag{89}$$

$$y' = y \tag{90}$$

$$z' = z \tag{91}$$

$$t' = \frac{t + \frac{V}{C^2}x}{\sqrt{1 + \frac{V^2}{C^2}}} \tag{92}$$

$$C^2 = \frac{1}{\varepsilon\mu} \tag{93}$$

Thus,

$$E = -mC^2 = -\frac{m_0 C^2}{\sqrt{1+\frac{V^2}{C^2}}} = -m_0 C^2 + \frac{1}{2}m_0 V^2 \cdots\cdots \tag{94}$$

$$\mathbf{p} = m\mathbf{V} = \frac{m_0}{\sqrt{1+\frac{V^2}{C^2}}}\mathbf{V} = m_0 \mathbf{V} \cdots\cdots \tag{95}$$

$$E^2 = -p^2 C^2 + m_0^2 C^4 \tag{96}$$

The other transformation is

$$x' = \frac{x + Vt}{\sqrt{1+\frac{V^2}{C^2}}} \tag{97}$$

$$y' = y \tag{98}$$

$$z' = z \tag{99}$$

$$t' = \frac{t - \frac{V}{C^2}x}{\sqrt{1+\frac{V^2}{C^2}}} \tag{100}$$

and

$$E = -mC^2 = -\frac{m_0 C^2}{\sqrt{1+\frac{V^2}{C^2}}} = -m_0 C^2 + \frac{1}{2}m_0 V^2 \cdots\cdots \tag{101}$$

$$\mathbf{p} = -m\mathbf{V} = -\frac{m_0}{\sqrt{1+\frac{V^2}{C^2}}}\mathbf{V} = -m_0 \mathbf{V} \cdots\cdots \tag{102}$$

$$E^2 = -p^2 C^2 + m_0^2 C^4 \tag{103}$$